\documentclass[aps,prl,preprint,groupedaddress,nofootinbib,showpacs]{revtex4-1}
\usepackage{hyperref}
\usepackage{mathrsfs}
\usepackage{bbm}
\usepackage{amsfonts}
\usepackage{amsmath}
\usepackage{graphicx}
\usepackage{subfig}
\usepackage{feynmp}

\newcommand{\void}[1]{}

\def\be{\begin{equation}}
\def\ee{\end{equation}}
\def\bea{\begin{eqnarray}}
\def\eea{\end{eqnarray}}
\def\lmd{\lambda}
\def\la{\langle}
\def\ra{\rangle}

\newcommand{\td}[1]{\tilde{#1}}

\newcommand{\bary}{\begin{array}}
\newcommand{\eary}{\end{array}}

\def\nb{\nonumber}

\begin{document}

    \title{Ward identity implied recursion relations in Yang-Mills theory}

\author{
Gang~Chen,$^{1}$
}
{\affiliation{
$^{1}$Department of Physics, Nanjing University\\
22 Hankou Road, Nanjing 210098, China
}

\hspace{1cm}
\begin{abstract}
The Ward identity in gauge theory constrains the behavior of the amplitudes. We discuss the Ward identity for amplitudes with a pair of shifted lines with complex momenta. This will induce a recursion relation identical to BCFW recursion relations at the finite poles of the complexified amplitudes. Furthermore, according to the Ward identity, it is also possible to transform  the boundary term into a simple form, which can be obtained by a new recursion relation. For the amplitude with one off-shell line in pure Yang-Mills theory, we find this technique is effective  for  obtaining the amplitude even when there are boundary contributions. 
\end{abstract}


\date{\today}
\maketitle
 Along with the breakthroughs in the spinor helicity technique and the formalism in twistor space-time\cite{Parke:1986gb,Xu:1986xb,Berends:1987me,Kosower,Dixon1,Witten1}, BCFW \cite{Britto:2004nj,Britto:2004nc,Britto:2004ap} recursion relation was found for tree level amplitudes in gauge theory. It was proven in \cite{Britto:2005fq} according to the singularity properties of the tree-level on-shell amplitudes. Rendering BCFW is an important technique in the analysis and calculation of  the amplitudes in various quantum fields theories even with massive fields \cite{Badger1,Ozeren,Schwinn,Chen1,Chen2}.  Recently, the BCFW technique is generalized to  the rational parts of the loop amplitudes\cite{Dixon4,Bern,Bern1}. And for the N=4 planar super Yang-Mills, N. \cite{Nima} have constructed the recursion relations for all loop integrands of the amplitudes. 

In BCFW formalism, the vanishing of the boundary term is necessary for the application of the  recursion relation. However, there are still various kinds of amplitudes which do have boundary terms. In this letter, we will use the Ward identity with  momenta shifted properly to determine the amplitudes in gauge theory.  For the poles at finite position, the residues are the same as those in BCFW recursion relation.  Moreover, this Ward identity will lead to  a new form for the boundary terms, which can be obtained by another recursion relation.  As an application, we will focus on the vector off-shell current which is the amplitude with one external line amputated and its momenta extended to off-shell. The method can also be extended to the tensor off-shell currents with several external off-shell lines. Our method is particularly useful for the case that the on-shell lines are of different  helicity structures. In this sense, our technique is complementary to the off-shell current recursion relation presented in \cite{Berends:1987me,Kosower}.

\textbf{Complexified Ward identity} In an  interacting gauge theory,  the Ward identity is a consequence of  the gauge current conservation.  As a result, the tensor currents  $\mathcal{A}^{\mu_1\mu_2\cdots \mu_m}$ vanish when contracted with all the momenta of the external off-shell lines
\be
k^1_{\mu_1} k^2_{\mu_2}\cdots k^m_{\mu_m} \mathcal{A}^{\mu_1\mu_2\cdots \mu_m}=0,
\ee  
where $m$ is the number of the external off-shell lines and $k^i_{\mu_i}$ denote the their momenta. Tensor current  is defined to be the amplitude with the propagators of the external off shell lines removed. The definition here has a subtle difference with the one in \cite{Dixon1}.  As long as the current conservation is not broken by  quantum corrections, the identity holds at each level of the perturbative expansion.  Furthermore, at tree level, the Ward identity holds even when the momenta are complexified with the on-shell condition untouched.   Hence it contains much more information about the off-shell currents than just the current conservation.  There is a simple proof for the complexified Ward identity directly according to the Feynman rules in Lorentz-Feynman gauge, where the tree level 2, 3 and 4 point vertices take the forms
\bea
V^2_{\mu\nu}&=&-i{\eta_{\mu\nu}\over k^2}\nb\\
V^3_{\mu_1\mu_2\mu_3}&=&{i\over\sqrt 2}\left(\eta_{\mu_1\mu_2}(k_1-k_2)_{\mu_3}+\eta_{\mu_2\mu_3}(k_2-k_3)_{\mu_1}+\eta_{\mu_3\mu_1}(k_3-k_1)_{\mu_2}\right)\nb\\
V^4_{\mu_1\mu_2\mu_3\mu_4}&=&{i\over 2}\left(2\eta_{\mu_1\mu_3}\eta_{\mu_2\mu_4}-\eta_{\mu_1\mu_2}\eta_{\mu_3\mu_4}-\eta_{\mu_1\mu_4}\eta_{\mu_2\mu_3}\right), 
\eea
there are three kinds of cancelations among the Feynman diagrams. Firstly according to inductive assumption about the conservation of the off-shell currents with fewer external on-shell lines,  the three-point vertices attached with the external off-shell line, when multiplied with the external off-shell momentum, can be simplified as $V^3_{\mu_1\mu_2\mu_3}k_3^{\mu_3}={i\over\sqrt 2}\eta_{\mu_1\mu_2}(k^2_2-k^2_1)$. Secondly, the summation of diagrams as shown in Fig.1 will vanish under the inductive assumption.
\begin{figure}[htb]
\centering
\includegraphics{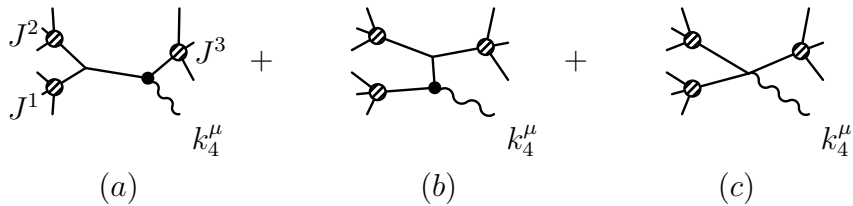}
\caption{In the diagrams we use $\bullet$  to denote that, in the three point vertex connecting with external momentum line, only the term ${i\over\sqrt 2}\eta_{\mu_{12}\mu_3}(-k^2_{12})$ contributes to (a) while ${i\over\sqrt 2}\eta_{\mu_1\mu_{23}}(k^2_{23})$ contributes for (b).}
\end{figure}
Finally,  the sum of diagrams in Fig.2 also vanish.  In a forthcoming article, we will prove in detail that all the  diagrams can be arranged into classes in Fig. 1 or  in Fig. 2.  Using  the second and third cancelations, the sum of contributions of the diagrams in each class is zero. Hence the current conservation holds under the recursive assumption. To complete the proof, we only need to verify the conservation of currents with only two on-shell lines, which can be  verified directly according to Feynman rules. 
\begin{figure}[htb]
\centering
\includegraphics{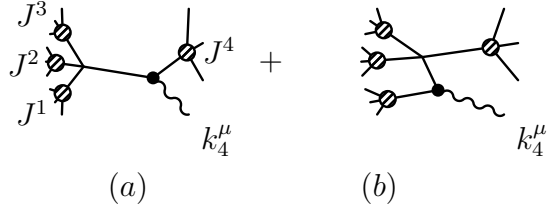}
\caption{Similarly as Fig.1, in the three point vertex connecting with external momentum line, only the term ${i\over\sqrt 2}\eta_{\mu_{123}\mu_3}(-k^2_{123})$ counts for (a) while ${i\over\sqrt 2}\eta_{\mu_1\mu_{234}}(k^2_{234})$ counts for (b).}
\end{figure}

In this article,  we always shift the momenta of a pair of lines $i,v$, where $i$ is one on-shell line with external state $\epsilon_i$ and $v$ is the off-shell line.  The momentum shift is chosen such as to keep the momentum conserved and the $i$-line on-shell.  For example,  we can shift the momenta as $\hat p_i=p_i-z\eta_i$ and  $\hat p_v=p_v-z\eta_i$, where $z$ is an arbitrary complex parameter and $\eta_i$ is a four-vector which satisfies $\eta_i\cdot \epsilon_i?0$. Then we get a complexified form of the Ward identity.  Acting with the first order derivative  with respect to $z$ on the complexified Ward identity, we obtain
\be\label{recz}
\mathcal{\hat A}(z)_\mu \eta^\mu_i=-\hat p_v^\mu {d\mathcal{\hat A}(z)_\mu\over dz},
\ee
where $z \eta_i=\hat p_i-p_i$ and $\mathcal{\hat A}(z)_\mu$ denotes the complexified vector off-shell current.
For convenience, in practical calculations,  we shift the momenta as $\lmd_i\rightarrow\lmd_i-z\lmd_v, \td\lmd_v\rightarrow\td\lmd_v+z\td\lmd_i$ for currents with $+$ helicity state $\epsilon_i^+={\mu\td\lmd_i\over \langle\mu,\lmd_i\rangle}$ in $i$-line, and $\eta_i=\lmd_v\td\lmd_i$. To avoid the unphysical pole from the external wave functions, we should choose the reference spinor as $\mu=\lmd_v$.  Similarly, for the negative states $\epsilon_i^-={\lmd_i\td\mu\over \langle\td\lmd_i,\td\mu\rangle}$, we can shift its  momentum  as $\td\lmd_i\rightarrow\td\lmd_i-z\td\lmd_v, \lmd_v\rightarrow\lmd_v+z\lmd_i$. For the same reason, the reference spinor is taken as $\td\mu=\td\lmd_v$. 

In pure Yang-Mills theory, we can expand $\mathcal{\hat A}(z)_\mu$ with respect to $z$ as 
\be\label{ExpA}
\mathcal{\hat A}(z)_\mu= A^1_\mu z +A^0_\mu+A^{-1_a}_\mu {1\over z-a}+A^{-1_b}_\mu {1\over z-b}+\cdots
\ee

According to (\ref{recz}),  it is easy to find the term $A^0_\mu$ should not contribute to the amplitude.  
Comparing (\ref{recz}) with (\ref{ExpA}),  the  boundary terms should be 
\be\label{boundequ}
A^0\cdot \eta_i=-{A^1\cdot p_v } .
\ee
Therefore $A^0$, which is hard to obtain directly, can be transformed to the $A^1$ which be obtained by a new recursion relation.

\textbf{Off-shell vector current from the Ward identity}
Now we apply our technique to the off-shell vector currents. Without loss of generality, we choose the shifted on-shell line to be of $+$ helicity. Under the choice of the momentum shift and the  reference spinor described above, the overall behavior of the off-shell currents are $z^1$ when $z\rightarrow \infty$. 
According to eq. (\ref{recz}), we get
\bea\label{OffAz}
\eta^\mu_i \hat A_\mu&=& -\hat p_v^\mu A^1_\mu + \sum_m\sum_h \hat p_v^\mu {A_L^h(z_m)(A_R^{\td h})_\mu(z_m)\over 2P_m \cdot \eta_i (z-z_m)^2} \nb\\
&=& -\hat p_v^\mu A^1_\mu + \sum_m\sum_h  {A_L^h(z_m)(A_R^{\td h})_\mu(z_m) \eta^\mu_i \over 2P_m \cdot \eta_i (z-z_m)},
\eea
where the we choose the off-shell line in $A_R^{\td h}$.
Since the factor $A_L^{k_m}(z_m)$ vanishes, we only need to take the summation over $(h,\td h)\in \{(+,-),(-,+),(r,k)\}$\cite{Feng}. Then the current's projection on $\eta_i$ can be obtained by setting $z=0$ in (\ref{OffAz})
\bea\label{OffA}
\eta^\mu_i A_\mu
&=& -p_v^\mu A^1_\mu + \sum_m\sum_h  {A_L^h(z_m)(A_R^{\td h})_\mu(z_m)\eta^\mu_i \over 2P_m\cdot  \eta_i (-z_m)}.
\eea

Using the  complexified Ward identity, we can hence transform the unknown term $A^0$ into $A^1$ which can also be obtained by a new recursion relation.  In fact when we choose the momentum shift and the gauge of the external on-shell states as discussed above, the wave function does not depend on $z$. Moreover, the four point vertices do not contain the momentum factor,  they are also independent of $z$. The $z$ dependent terms only come from the three point vertices and propagators in the complex lines from external line $i$ to $v$.  $A^1_\mu =({d\mathcal{\hat A}(z)_\mu \over dz})^0$ then gets two kinds of contributions as shown in Fig. 3.   
\begin{figure}[htb]
\centering
\includegraphics{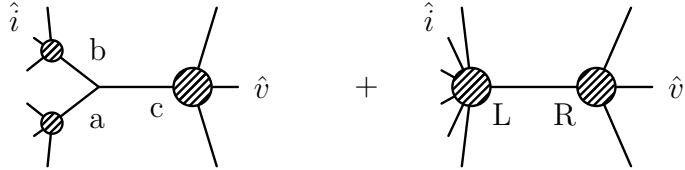}
\caption{The Feynman diagrams which will contribute to the boundary terms}
\end{figure}
When acting  with ${d\over dz}$ on the propagators and extracting the zeroth order terms in $z$,   we get 
\be
i( A_L^{\nu_L})^1{g_{\nu_L\nu_R}\over 2P_m\cdot\eta_i } (A_R^{\nu_R \mu})^1. 
\ee 
When acting with ${d\over dz}$ on the three point vertices, we get
\be
 {1\over \sqrt{2}} {A_a\cdot (A_b)^1~ \eta_i\cdot (A_c^\mu)^1\over 2p_b\cdot\eta_i~ 2p_c\cdot \eta_i ~p_a^2}-\sqrt{2} {(A_b)^1 \cdot(A_c^\mu)^1~ \eta_i\cdot A_a\over 2p_b\cdot\eta_i~ 2p_c\cdot \eta_i ~p_a^2}. 
 \ee
 Finally, we take the summation of all the complexified three-point vertices and the propagators 
\be
(A^{\mu})^1=\sum_{a,b,c}\left({1\over \sqrt{2}} {A_a\cdot (A_b)^1~ \eta_i\cdot (A_c^\mu)^1\over 2p_b\cdot\eta_i~ 2p_c\cdot \eta_i ~p_a^2} -\sqrt{2} {(A_b)^1 \cdot (A_c^\mu)^1~ \eta_i\cdot A_a\over 2p_b\cdot\eta_i~ 2p_c\cdot \eta_i ~p_a^2}\right)+ \sum_m i{(A_L)^1\cdot (A_R^\mu)^1\over 2p_m\cdot\eta_i}.
\ee

To complete the recursion relation, we need to know the coefficients of order $z$   in the tensor off-shell  currents $(\hat A^{\nu\mu})^1$.  It is similar to the $(A^\mu)^1$,
\bea
(A^{\nu\mu})^1&=&{1\over \sqrt{2}} {A_a\cdot (A_b^\nu)^1~ \eta_i\cdot (A_c^\mu)^1\over 2p_b\cdot\eta_i~ 2p_c\cdot \eta_i ~p_a^2} -\sqrt{2} {(A_b^\nu)^1 \cdot (A_c^\mu)^1~ \eta_i\cdot A_a\over 2p_b\cdot\eta_i~ 2p_c\cdot \eta_i ~p_a^2}
+ i{(A_L^\nu)^1\cdot (A_R^\mu)^1\over 2p_m\cdot\eta_i}.
\eea

In (\ref{OffA}), there are new non-vanishing  objects which can be taken as the off-shell amplitudes with one external states of the on-shell lines replaced by its momentum.  On proceeding several recursive steps,  we get  a general form $\mathcal{A}^{\mu}(\cdots, k_{i_1},\cdots, k_{i_j},\cdots,k_{i_N})$, where we omit on-shell states with $N$ denoting the total number of the replaced lines.   

Inevitably, we will need to shift the momentum of such line together with the off-shell line.  The boundary term then can not be obtained as above.  Under the momentum  shift,  $\lmd_i\rightarrow\lmd_i-z\lmd_v, \td\lmd_v\rightarrow\td\lmd_v+z\td\lmd_i$, the vector currents are of the form $\hat k^\nu_i \hat A_{\nu\mu}$.  The shifted momentum contains a factor proportional  to $z$. This will lead to $(\hat A^\mu)^0$  contributing to the boundary term. Luckily, we only need to know the amplitude when $z\rightarrow 0$.  As it is obvious that $\hat k^\nu_i \hat A_{\nu\mu}|_{z=0}= k^\nu_i \hat A_{\nu\mu}|_{z=0}$, we only need to consider  $ k^\nu_i \hat A_{\nu\mu}|_{z=0}$.

There is a similar identity for calculating this kind of currents
\be\label{newId}
{ k^\nu_i \hat A_{\nu\mu} \hat K^\mu\over [\td\lmd_m,\td\lmd_v]}=0,
\ee
which can be deduced from the Ward identity,
\bea
{\hat  k^\nu_i \hat A_{\nu\mu} \hat K^\mu \over [\td\lmd_m,\td\lmd_v]}&=&0\nb\\
{\eta^\nu_i \hat A_{\nu\mu} \hat K^\mu\over [\td\lmd_m,\td\lmd_v]}&=&\epsilon^\nu_i \hat A_{\nu\mu} \hat K^\mu=0.
\eea
From (\ref{newId}), we  get a similar recursion relation for $k^\nu_i  A_{\nu\mu} \eta_i^\mu$.  Similarly, for another momentum shift $\td\lmd_i\rightarrow\td\lmd_i-z\td\lmd_v, \lmd_v\rightarrow\lmd_v+z\lmd_i$, we  get a recursion relation for $k^\nu_i  A_{\nu\mu} \td\eta_i^\mu$. 
Now we have successfully expressed  the boundary term in the amplitude as terms composed of  the off-shell amplitudes with fewer external lines.  The component of the off-shell vector current in the momentum direction vanishes according to the Ward identity.  To get the full off-shell vector current, we hence need to project onto three linearly independent directions, none of which is parallel with the momentum.   We can obtain all of them using the same procedure above. 

Without loss of generality, we choose the lines $(i,j,k)$ with helicity $(+,+,-)$ respectively.  In the same way, we  obtain $\eta_j\cdot \ A$ and $\td\eta_k \cdot A$ where $\eta_j=\lmd_v\td\lmd_j$ and $\td\eta_k=\lmd_k\td\lmd_v$. It is convenient to write the vector current as $A^\mu=x_1\eta^\mu_i+x_2 \eta^\mu_j +x_3 \td\eta^\mu_k+x_4 K^\mu_v$.  We then determine the off-shell vector currents by solving the following four equations
\bea
e_{ik}x_3+e_{iv}x_4&=&\eta_i\cdot A\nb\\
e_{jk}x_3+e_{jv}x_4&=&\eta_j\cdot A\nb\\
e_{ki}x_1+e_{kj}x_2+e_{kv}x_4&=&\eta_k\cdot A\nb\\
e_{vi}x_1+e_{vj}x_2+e_{vk}x_3+e_{vv}x_4&=&0,
\eea
where $e_{ij}$ are the inner products of the basis vectors and the lower indices denote the corresponding basis. 
Hence, to get the full vector current $A^\mu$, we only need to choose three different kinds of momentum shift, by choosing  different external lines or different shifts for one line, such that the shifted momenta are linearly independent.  This is possible  for any vector currents with three external on-shell lines.

There is a direct verification for the boundary term equation (\ref{boundequ}) for three line off-shell amplitudes.  For this, purpose we take $A(1^+,2^-,\mu)$ as an example. We choose $1$ and $3$ as the shifting lines.  The momentum shift and the reference spinors  for the external lines are chosen as discussed above.  It is easy to show that  $A^0\cdot \eta_1=-A^1\cdot k_3$ 
\bea
A^0\cdot \eta_1&=&{i\over \sqrt{2}}\left(-\epsilon^{+}_1\cdot \epsilon^{-}_2 k_2\cdot \eta_1+2k_2\cdot \epsilon^{+}_1 \epsilon^{-}_2\cdot\eta_1\right) ={i\over \sqrt{2}} \epsilon^{+}_1\cdot \epsilon^{-}_2 k_2\cdot \eta_1\nb\\
-A^1\cdot k_3&=&{i\over \sqrt{2}}\left(\epsilon^{+}_1\cdot \epsilon^{-}_2 k_3\cdot \eta_1-2k_3\cdot \epsilon^{+}_1 \epsilon^{-}_2\cdot\eta_1 \right)={i\over \sqrt{2}} \epsilon^{+}_1\cdot \epsilon^{-}_2 k_2\cdot \eta_1.
\eea

Another simple example is vector currents with three on-shell lines.  For concreteness, we choose the current to be $A(1^+,2^+,3^-, \mu_4)$.  As stated above, the reference momenta and spinors are taken as $k_v=\lmd_v\td\lmd_v$ and $\mu_1=\mu_2=\lmd_v, \td\mu_3=\td\lmd_v$.  The shifting momenta  are $\eta_1=\lmd_v\td\lmd_1,\eta_2=\lmd_v\td\lmd_2$ and $\td\eta_3=\td\lmd_v \lmd_3$ for the states $\epsilon^{+}_1={\lmd_v\td\lmd_1 \over \la\lmd_v,\lmd_1\ra},\epsilon^{+}_2={\lmd_v\td\lmd_2 \over \la\lmd_v, \lmd_2\ra} $ and $\epsilon^{-}_3={\lmd_3\td\lmd_v \over [\td\lmd_3, \td\lmd_v]}$ respectively.  Then we get the  components of the vector off-shell currents $A\cdot \eta_1, A\cdot \eta_2$ and $A \cdot  \td\eta_3$.  We will compare the results in our methods with those in usual Feynman rules for $A\cdot \eta_1$.  The other two are similar. In Feynman rules the three-line vector currents can be writen as 
\bea\label{aterm}
-{1\over \sqrt 2}{\eta^\mu_1\cdot A(\mu,2^+,3^-) k_{4}\cdot \epsilon_1\over p_{23}^2}
+{A(1^+, 2^+,\nu_1)  g^{\nu_1\nu_2}A(\nu_2,3^-,\mu)\eta_1^\mu \over p_{12}^2}
\eea
Using our methods, we can get one component of the off-shell current
\bea\label{ex1}
A\cdot \eta_1&=&-p_4\cdot A^1+{A(1^+, 2^+,m^-) A(m^+,3^-,\mu)\eta_1^\mu \over2 P_m\cdot \eta_1 (-z_m)}\nb\\
&=&-{1\over \sqrt 2}{\eta^\mu_1A(\mu,2^+,3^-) \epsilon_1\cdot p_4\over p_{23}^2}+{A(1^+, 2^+,m^-) A(m^+,3^-,\mu)\eta_1^\mu \over p_{12}^2}
\eea
The second term in the above equation  can be transformed as follows 
\bea
&&{A^{z_m}(1^+, 2^+,m^-) A^{z_m}(m^+,3^-,\mu)\eta_1^\mu \over p_{12}^2}={A^{z_m}(1^+, 2^+,\nu_1) g^{\nu_1\nu_2}A^{z_m}(\nu_2,3^-,\mu)\eta_1^\mu \over p_{12}^2}\nb\\
&=&{A(1^+, 2^+,\nu_1)  g^{\nu_1\nu_2}\left(A(\nu_2,3^-,\mu)\eta_1^\mu+{i\over \sqrt{2}}\left(z_m (\epsilon^{+}_2)_{v_2} (-\eta_1)_{\mu}\eta_1^\mu+2z_m (\eta_1)_{v2}\eta_1\cdot \epsilon^{-}_3 \right)\right) \over p_{12}^2}\nb\\
&=&{A(1^+, 2^+,\nu_1)  g^{\nu_1\nu_2}A(\nu_2,3^-,\mu)\eta_1^\mu \over p_{12}^2}\nb\\
\eea
showing the equality of (\ref{aterm}) and (\ref{ex1}). 

In summary, using the shifted Ward identity, we propose a new form for the boundary term for the BCFW shifted amplitude or off-shell vector currents. In this way, the boundary terms can be obtained by a new recursion relation, which is exactly the usual BCFW recursion relation when reduced to single poles. We apply our technique to the off-shell currents with on-shell lines  with different helicity structures. It is easy to see that our technique is more efficient for the currents of general helicity structures for the on-shell lines, complimenting the existent off-shell recursion relation. First of all, the number of the effective diagrams is small.  For finite poles contribution, only the propagator along the complex line contributes; while for the  boundary terms  both the propagator and half parts in the three-point vertex  contribute to the vector currents. Four-point gluon interaction needs no consideration.    
In proceeding the recursion relation, there will be new off-shell currents with some of the on-shell states  replaced by their momenta.  Such new objects can also be obtained in our techniques.

Although we focus on the one-line off-shell vector currents in gauge theory, the technique from complex Ward identity can be generalize to theories with gauge symmetry spontaneously broken as well as to tensor currents with several off-shell lines.  The current with two off-shell line is important for constructing one-loop amplitudes. Another extension is to study the amplitude at one loop level according to the loop level Ward identity. However,  the complex Ward identity does not present itself at the loop level, it warrants further study.  

\textbf{Acknowledgement} We thank Edna Cheung, Jens Fjelstad, Konstantin G. Savvidy for helpful discussions.  This work is funded by the Priority Academic Program Development of Jiangsu Higher Education Institutions (PAPD), NSFC grant No.~10775067, Research Links Programme of Swedish Research Council under contract No.~348-2008-6049, the Chinese Central Government's 985 Project grants for Nanjing University, the China Science Postdoc grant no. 020400383. the postdoc grants of Nanjing University 0201003020


\begin{thebibliography}{99}
\bibitem{Parke:1986gb}
  S.~J.~Parke and T.~R.~Taylor,
  Phys.\ Rev.\ Lett.\  {\bf 56}, 2459 (1986).

\bibitem{Xu:1986xb}
  Z.~Xu, D.~H.~Zhang and L.~Chang,
  Nucl.\ Phys.\  B {\bf 291}, 392 (1987).

\bibitem{Berends:1987me}
  F.~A.~Berends and W.~T.~Giele,
  Nucl.\ Phys.\  B {\bf 306}, 759 (1988).
  
\bibitem{Kosower} D. A. Kosower, Nucl. Phys. B\textbf{335}:23 (1990).

\bibitem{Dixon1} L. J. Dixon,  Bouler TASI \textbf{95}, 539-584.

\bibitem{Witten1} E. Witten, Commun. Math. Phys {\bf 252}, 189-258.


\bibitem{Britto:2004nj}
  R.~Britto, F.~Cachazo and B.~Feng,
  Phys.\ Rev.\ D {\bf 71}, 025012 (2005)

\bibitem{Britto:2004nc}
 R.~Britto, F.~Cachazo and B.~Feng, Nucl. Phys. B \textbf{725}, 275-305 (2005).


\bibitem{Britto:2004ap}
 R.~Britto, F.~Cachazo and B.~Feng, Nucl. Phys. B \textbf{715}, 499-522 (2005).


\bibitem{Britto:2005fq}
R.~Britto, F.~Cachazo, B.~Feng and E.~Witten, Phys. Rev. Lett. \textbf{94}, 181602 (2005).

\bibitem{Badger1} S. D. Badger, E. W. Glover, V. V. Khoze and P. Svrcek, JHEP \textbf{0507}, 025 (2005).


\bibitem{Ozeren} K. J. Ozeren and W.  J. Stirling, Eur. Phys. J. C \textbf{48}, 159-168 (2006).

\bibitem{Schwinn} C. Schwinn, Phys. Rev. D \textbf{78}, 085030 (2008).

\bibitem{Chen1} G. Chen, K. G. Savvidy, Eur. Phys. J. C \textbf{72}, 3, 1952 (2012).
\bibitem{Chen2} G. Chen, Phys.Rev. D83 (2011) 125005.

\bibitem{Dixon4} Z. Bern, V. Del Duca, L. J. Dixon, and D. A. Kosower, Phys. Rev. D \textbf{71}, 045006 (2005).

\bibitem{Bern} Z. Bern, N. E. J. Bjerrum-Bohr, David C. Dunbar and Harald Ita, JHEP 0511 (2005) 027.

\bibitem{Bern1} Carola F. Berger, Zvi Bern, Lance J. Dixon, Darren Forde, David A. Kosower, Phys.Rev. D74 (2006) 036009.

\bibitem{Nima} Nima Arkani-Hamed, Jacob L. Bourjaily, Freddy Cachazo, Simon Caron-Huot, Jaroslav Trnka, JHEP 1101 (2011) 041.

\bibitem{Feng} B. Feng, Z.B. Zhang, JHEP 1112 (2011) 057.




\end{thebibliography}
\end{document}